\documentclass[letterpaper,prb,twocolumn,showpacs]{revtex4}
\usepackage{amsmath}
\usepackage{amssymb}
\usepackage[dvips]{graphicx}
\usepackage[dvips]{color}
\usepackage{bbold}
\usepackage{subfigure}

\begin{document}

\title{Magnetization Dissipation in Ferromagnets from Scattering Theory}
\author{Arne Brataas}
\email{Arne.Brataas@ntnu.no}
\affiliation{Department of Physics, Norwegian University of Science and Technology,
NO-7491 Trondheim, Norway}
\author{Yaroslav Tserkovnyak}
\affiliation{Department of Physics and Astronomy, University of California, Los Angeles,
California 90095, USA}
\author{Gerrit E. W. Bauer}
\affiliation{Institute for Materials Research, Tohoku University, Sendai 980-8577, Japan}
\affiliation{Kavli Institute of NanoScience, Delft University of Technology, Lorentzweg
1, 2628 CJ Delft, The Netherlands}

\begin{abstract}
The magnetization dynamics of ferromagnets are often formulated in terms of
the Landau-Lifshitz-Gilbert (LLG) equation. The reactive part of this
equation describes the response of the magnetization in terms of effective
fields, whereas the dissipative part is parameterized by the Gilbert damping
tensor. We formulate a scattering theory for the magnetization dynamics and
map this description on the linearized LLG equation by attaching electric
contacts to the ferromagnet. The reactive part can then be expressed in
terms of the static scattering matrix. The dissipative contribution to the
low-frequency magnetization dynamics can be described as an adiabatic energy
pumping process to the electronic subsystem by the time-dependent
magnetization. The Gilbert damping tensor depends on the time derivative of
the scattering matrix as a function of the magnetization direction. By the
fluctuation-dissipation theorem, the fluctuations of the effective fields
can also be formulated in terms of the quasistatic scattering matrix. The
theory is formulated for general magnetization textures and worked out for
monodomain precessions and domain wall motions. We prove that the Gilbert
damping from scattering theory is identical to the result obtained by the
Kubo formalism.
\end{abstract}

\pacs{75.40.Gb,76.60.Es,72.25.Mk}
\keywords{}
\maketitle

\section{Introduction}

\label{introduction}

Ferromagnets develop a spontaneous magnetization below the Curie
temperature. The long-wavelength modulations of the magnetization direction
consist of spin waves, the low-lying elementary excitations (Goldstone
modes) of the ordered state. When the thermal energy is much smaller than
the microscopic exchange energy, the magnetization dynamics can be
phenomenologically expressed in a generalized Landau-Lifshitz-Gilbert (LLG)
form: 
\begin{align}
\mathbf{\dot{m}}(\mathbf{r},t)& =-\gamma \mathbf{m}(\mathbf{r},t)\times %
\left[ \mathbf{H}_{\mathrm{eff}}(\mathbf{r},t)+\mathbf{h}(\mathbf{r},t)%
\right] +  \notag \\
& \mathbf{m}(\mathbf{r},t)\times \int d\mathbf{r}^{\prime }\left[ \tilde{%
\alpha}\left[ \mathbf{m}\right] (\mathbf{r,r}^{\prime })\mathbf{\dot{m}}(%
\mathbf{r}^{\prime },t)\right] ,  \label{LLG}
\end{align}%
where the magnetization texture is described by $\mathbf{m}(\mathbf{r},t),$
the unit vector along the magnetization direction at position $\mathbf{r}$
and time $t$, $\mathbf{\dot{m}}(\mathbf{r},t)=\partial \mathbf{m}(\mathbf{r}%
,t)/\partial t$, $\gamma =g\mu _{B}/\hbar $ is the gyromagnetic ratio in
terms of the $g$-factor ($\approx 2$ for free electrons) and the Bohr
magneton $\mu _{B}$. The Gilbert damping $\tilde{\alpha}$ is a nonlocal
symmetric $3\times 3$ tensor that is a functional of $\mathbf{m}$. The
Gilbert damping tensor is commonly approximated to be diagonal and isotropic
(i), local (l), and independent of the magnetization $\mathbf{m}$, with
diagonal elements 
\begin{equation}
\alpha _{\mathrm{il}}(\mathbf{r},\mathbf{r}^{\prime })=\alpha \delta (%
\mathbf{r}-\mathbf{r}^{\prime }).
\end{equation}%
The linearized version of the LLG equation for small-amplitude excitations
has been derived microscopically.\cite{Heinrich:pss67} It has been used very
successfully to describe the measured response of ferromagnetic bulk
materials and thin films in terms of a small number of adjustable,
material-specific parameters. The experiment of choice is ferromagnetic
resonance (FMR), which probes the small-amplitude coherent precession of the
magnet.\cite{Bland:book05} The Gilbert damping model in the local and
time-independent approximation has important ramifications, such as a linear
dependence of the FMR line width on resonance frequency, that have been
frequently found to be correct. The damping constant is technologically
important since it governs the switching rate of ferromagnets driven by
external magnetic fields or electric currents.\cite{Stiles:top06} In
spatially dependent magnetization textures, the nonlocal character of the
damping can be significant as well.\cite{Foros:prb08,Zhang:prl09,Wong:prb09}
Motivated by the belief that the Gilbert damping constant is an important
material property, we set out here to understand its physical origins from
first principles. We focus on the well studied and technologically important
itinerant ferromagnets, although the formalism can be used in principle for
any magnetic system.

The reactive dynamics within the LLG Eq. (\ref{LLG}) is described by the
thermodynamic potential $\Omega \lbrack \mathbf{M}]$ as a functional of the
magnetization. The effective magnetic field $\mathbf{H}_{\mathrm{eff}}[%
\mathbf{M}](\mathbf{r})\equiv -\delta \Omega /\delta \mathbf{M}(\mathbf{r})$
is the functional derivative with respect to the local magnetization $%
\mathbf{M}(\mathbf{r})=M_{s}\mathbf{m}(r)$, including the external magnetic
field $\mathbf{H}_{\mathrm{ext}}$, the magnetic dipolar field $\mathbf{H}_{%
\mathrm{d}}$, the texture-dependent exchange energy, and crystal field
anisotropies. $M_{s}$ is the saturation magnetization density. Thermal
fluctuations can be included by a stochastic magnetic field $\mathbf{h}(%
\mathbf{r},t)$ with zero time average, $\left \langle \mathbf{h}\right \rangle
=0$, and white-noise correlation:\cite{Brown:pr63} 
\begin{equation}
\left \langle h_{i}(\mathbf{r},t)h_{j}(\mathbf{r}^{\prime },t^{\prime
})\right \rangle =\frac{2k_{B}T}{\gamma M_{s}}\tilde{\alpha}_{ij}\left[ 
\mathbf{m}\right] (\mathbf{r},\mathbf{r}^{\prime })\delta (t-t^{\prime }),
\label{LLG_FDT}
\end{equation}%
where $M_{s}$ is the magnetization, $i$ and $j$ are the Cartesian indices,
and $T$ is the temperature. This relation is a consequence of the
fluctuation-dissipation theorem (FDT) in the classical (Maxwell-Boltzmann)
limit.

\begin{figure}[tbp]
\includegraphics[scale=0.666]{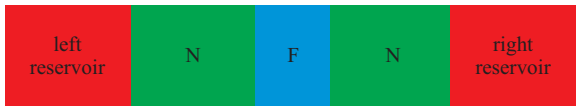}
\caption{Schematic picture of a ferromagnet (F) in contact with a thermal
bath (reservoirs) via metallic normal metal leads (N).}
\label{fig:scattering}
\end{figure}

The scattering ($S$-) matrix is defined in the space of the transport
channels that connect a scattering region (the sample) to real or fictitious
thermodynamic (left and right) reservoirs by electric contacts with leads
that are modeled as ideal wave guides. Scattering matrices are known to
describe transport properties, such as the giant magnetoresistance, spin
pumping, and current-induced magnetization dynamics in layered normal-metal
(N)$\mid $ferromagnet (F).\cite%
{Waintal:prb00,Brataas:prl00,Tserkovnyak:prl02} When the ferromagnet is part
of an open system as in Fig.\  \ref{fig:scattering}, also $\Omega $ can be
expressed in terms of the scattering matrix, which has been used to express
the non-local exchange coupling between ferromagnetic layers through
conducting spacers.\cite{Bruno:prb95} We will show here that the
scattering matrix description of the effective magnetic fields is valid even
when the system is closed, provided the dominant contribution comes from the
electronic band structure, scattering potential disorder, and spin-orbit
interaction.

Scattering theory can also be used to compute the Gilbert damping tensor $%
\tilde{\alpha}$ for magnetization dynamics.\cite{Brataas:prl08} The energy
loss rate of the scattering region can be expressed in terms of the
time-dependent $S$-matrix. To this end, the theory of adiabatic quantum
pumping has to be generalized to describe dissipation in a metallic
ferromagnet. The Gilbert damping tensor is found by evaluating the energy
pumping out of the ferromagnet and relating it to the energy loss that is
dictated by the LLG equation. In this way, it is proven that the Gilbert
phenomenology is valid beyond the linear response regime of small
magnetization amplitudes. The key approximation that is necessary to derive
Eq.~(\ref{LLG}) including $\tilde{\alpha}$ is the (adiabatic) assumption
that the ferromagnetic resonance frequency $\omega _{\mathrm{FMR}}$ that
characterizes the magnetization dynamics is small compared to internal
energy scale set by the exchange splitting $\Delta $ and spin-flip
relaxation rates $\tau _{s}$. The LLG phenomenology works well for
ferromagnets for which $\omega _{\mathrm{FMR}}\ll \Delta /\hbar $, which is
certainly the case for transition metal ferromagnets such as Fe and Co.

Gilbert damping in transition-metal ferromagnets is generally believed to
stem from the transfer of energy from the magnetic order parameter to the
itinerant quasiparticle continuum. This requires either magnetic disorder or
spin-orbit interactions in combination with impurity/phonon scattering.\cite%
{Bland:book05} Since the heat capacitance of the ferromagnet is dominated by
the lattice, the energy transferred to the quasiparticles will be dissipated
to the lattice as heat. Here we focus on the limit in which elastic
scattering dominates, such that the details of the heat transfer to the
lattice does not affect our results. Our approach formally breaks down in
sufficiently clean samples at high temperatures in which inelastic
electron-phonon scattering dominates. Nevertheless, quantitative insight can
be gained by our method even in that limit by modelling phonons by frozen
deformations.\cite{Liu:10}

In the present formulation, the heat generated by the magnetization dynamics
can escape only via the contacts to the electronic reservoirs. By computing
this heat current through the contacts we access the total dissipation rate.
Part of the heat and spin current that escapes the sample is due to spin
pumping that causes energy and momentum loss even for otherwise dissipation
less magnetization dynamics. This process is now well understood.\cite%
{Tserkovnyak:prl02} For sufficiently large samples, the spin pumping
contribution is overwhelmed by the dissipation in the bulk of the
ferromagnet. Both contributions can be separated by studying the heat
generation as a function of the length of a wire. In principle, a voltage
can be added to study dissipation in the presence of electric currents as in \onlinecite{Hals:prl09,Hals:epl10}, but
we concentrate here on a common and constant chemical potential in both
reservoirs.

Although it is not a necessity, results can be simplified by expanding the $%
S $-matrix to lowest order in the amplitude of the magnetization dynamics.
In this limit scattering theory and the Kubo linear response formalism for
the dissipation can be directly compared. We will demonstrate explicitly
that both approaches lead to identical results, which increases our
confidence in our method. The coupling to the reservoirs of large samples is
identified to play the same role as the infinitesimals in the Kubo approach
that guarantee causality.

Our formalism was introduced first in Ref. \onlinecite{Brataas:prl08} limited to the macrospin model and zero temperature. An extension to the friction associated with domain wall motion was given in Ref. \onlinecite{Hals:prl09}. Here we show how to
handle general magnetization textures and finite temperatures. Furthermore,
we offer an alternative route to derive the Gilbert damping in terms of the
scattering matrix from the thermal fluctuations of the effective field. We
also explain in more detail the relation of the present theory to spin and
charge pumping by magnetization textures.

Our paper is organized in the following way. In Section~\ref{Model}, we
introduce our microscopic model for the ferromagnet. In Section~\ref%
{Dissipation}, dissipation in the Landau-Lifshitz-Gilbert equation is
exposed. The scattering theory of magnetization dynamics is developed in
Sec.~\ref{Scattering}. We discuss the Kubo formalism for the time-dependent
magnetizations in Sec.~\ref{Kubo}, before concluding our article in Sec.~\ref%
{Conclusions}. The Appendices provide technical derivations of spin, charge,
and energy pumping in terms of the scattering matrix of the system.

\section{Model}

\label{Model}

Our approach rests on density-functional theory (DFT), which is widely and
successfully used to describe the electronic structure and magnetism in many
ferromagnets, including transition-metal ferromagnets and ferromagnetic
semiconductors.\cite{Kubler:2000} In the Kohn-Sham implementation of DFT,
noninteracting hypothetical particles experience an effective
exchange-correlation potential that leads to the same ground-state density
as the interacting many-electron system.\cite{Kohn:rmp99} A simple yet
successful scheme is the local-density approximation to the effective
potential. DFT theory can also handle time-dependent phenomena. We adopt
here the adiabatic local-density approximation (ALDA), \textit{i.e}. an
exchange-correlation potential that is time-dependent, but local in time and
space.\cite{Zangwill:prl80,Gross:prl85} As the name expresses, the ALDA is
valid when the parametric time-dependence of the problem is adiabatic with
respect to the electron time constants. Here we consider a magnetization
direction that varies slowly in both space and time. The ALDA should be
suited to treat magnetization dynamics, since the typical time scale ($t_{%
\text{FMR}}\sim 1/\left( 10~\text{GHz}\right) \sim 10^{-10}\text{s}$) is long compared to the that associated with the Fermi and exchange
energies, $1-10$ $\text{eV}$ leading to $\hbar /\Delta \sim 10^{-13}\text{s}$
in transition metal ferromagnets.

In the ALDA, the system is described by the time-dependent effective Schr%
\"{o}dinger equation%
\begin{equation}
\hat{H}_{\text{ALDA}}\Psi (\mathbf{r},t)=i\hbar \frac{\partial }{\partial t%
}\Psi (\mathbf{r},t),
\end{equation}%
where $\Psi (\mathbf{r},t)$ is the quasiparticle wave function at position $%
\mathbf{r}$ and time $t$. We consider a generic mean-field electronic
Hamiltonian that depends on the magnetization direction $\hat{H}_{\text{%
ALDA}}\left[ \mathbf{m}\right] $ and includes the periodic Hartree, exchange
and correlation potentials and relativistic corrections such as the
spin-orbit interaction. Impurity scattering including magnetic disorder is
also represented by $\hat{H}_{\text{ALDA}}.$ The magnetization $\mathbf{m}$
is allowed to vary in time and space. The total Hamiltonian depends
additionally on the Zeeman energy of the magnetization in external $\mathbf{H%
}_{\text{ext}}$ and dipolar $\mathbf{H}_{d}$ magnetic fields: 
\begin{equation}
\hat{H}=\hat{H}_{\text{ALDA}}[\mathbf{m}]-M_{s}\int dr\, \mathbf{m}\cdot
\left( \mathbf{H}_{\text{ext}}+\mathbf{H}_{d}\right) .  \label{Ham_tot}
\end{equation}

For this general Hamiltonian (\ref{Ham_tot}), our task is to deduce an
expression for the Gilbert damping tensor $\tilde{\alpha}$. To this end,
from the form of the Landau-Lifshitz-Gilbert equation (\ref{LLG_FDT}), it is
clear that we should seek an expansion in terms of the slow variations of
the magnetizations in time. Such an expansion is valid provided the
adiabatic magnetization precession frequency is much less than the exchange
splitting $\Delta $ or the spin-orbit energy which governs spin relaxation
of electrons. We discuss first dissipation in the LLG equation and
subsequently compare it with the expressions from scattering theory of
electron transport. This leads to a recipe to describe dissipation by first
principles. Finally, we discuss the connection to the Kubo linear response
formalism and prove that the two formulations are identical in linear
response.

\section{Dissipation and Landau-Lifshitz-Gilbert equation}

\label{Dissipation}

The energy dissipation can be obtained from the solution of the LLG Eq. (\ref%
{LLG}) as 
\begin{eqnarray}
\dot{E} &=&-M_{s}\int d\mathbf{r}\left[ \mathbf{\dot{m}}(\mathbf{r},t)\cdot 
\mathbf{H}_{\text{eff}}(\mathbf{r},t)\right]  \label{ediss} \\
&=&-\frac{M_{s}}{\gamma }\int d\mathbf{r}\int d\mathbf{r}^{\prime }\mathbf{%
\dot{m}(r})\cdot \tilde{\alpha}\left[ \mathbf{m}\right] \left( \mathbf{r},%
\mathbf{r}^{\prime }\right) \cdot \mathbf{\dot{m}(r}^{\prime }).
\end{eqnarray}%
The scattering theory of magnetization dissipation can be formulated for
arbitrary spatiotemporal magnetization textures. Much insight can be gained
for certain special cases. In small particles or high magnetic fields the
collective magnetization motion is approximately constant in space and the
\textquotedblleft macrospin\textquotedblright \ model is valid in which all
spatial dependences are disregarded. We will also consider special
magnetization textures with a dynamics characterized by a number of dynamic
(soft) collective coordinates $\xi _{a}(t)$ counted by $a$:\cite%
{Tretiakov:prl08,Clarke:prb08} 
\begin{equation}
\mathbf{m}(\mathbf{r},t)=\mathbf{m}_{\text{st}}(\mathbf{r};\left \{ \xi
_{a}(t)\right \} ),  \label{Collective_coordinate}
\end{equation}%
where $\mathbf{m}_{\text{st}}$ is the profile at $t\rightarrow -\infty .$
This representation has proven to be very effective in handling
magnetization dynamics of domain walls in ferromagnetic wires. The
description is approximate, but (for few variables) it becomes exact in
special limits, such as a transverse domain wall in wires below the Walker
breakdown (see below); it becomes arbitrarily accurate by increasing the
number of collective variables. The energy dissipation to lowest (quadratic)
order in the rate of change $\dot{\xi}_{a}$ of the collective coordinates is 
\begin{equation}
\dot{E}=-\sum_{ab}\tilde{\Gamma}_{ab}\dot{\xi}_{a}\dot{\xi}_{b},
\label{energy_dissipation_collective}
\end{equation}%
The (symmetric) dissipation tensor $\tilde{\Gamma}_{ab}$ reads 
\begin{widetext}
\begin{equation}
\tilde{\Gamma}_{ab}=\frac{M_{s}}{\gamma }\int d\mathbf{r}\int d\mathbf{r}
^{\prime }\frac{\partial \mathbf{m}_{\text{st}}\mathbf{(r})}{\partial \xi
_{a}}
\alpha \left[ \mathbf{m}\right] (\mathbf{r},\mathbf{r}^{\prime })
\cdot \frac{\partial \mathbf{m}_{\text{st}}\mathbf{(r}^{\prime })}{
\partial \xi _{b}}.  \label{diss_tensor}
\end{equation}
\end{widetext}The equation of motion of the collective coordinates under a
force%
\begin{equation}
\boldsymbol{\mathfrak{F}}=-\frac{\partial \Omega }{\partial \boldsymbol{\xi }%
}
\end{equation}%
are\cite{Tretiakov:prl08,Clarke:prb08} 
\begin{equation}
\tilde{\eta}\boldsymbol{\dot{\xi}}+\left[ \boldsymbol{\mathfrak{F}}+%
\boldsymbol{\mathfrak{f}}(t)\right] -\tilde{\Gamma}\boldsymbol{\dot{\xi}}=0,
\label{equation_collective}
\end{equation}%
introducing the antisymmetric and time-independent gyrotropic\ tensor: 
\begin{equation}
\tilde{\eta}_{ab}=\frac{M_{s}}{\gamma }\int d\mathbf{r\,}\mathbf{m}_{\text{%
st}}\mathbf{(r)}\cdot \left[ \frac{\partial \mathbf{m}_{\text{st}}\mathbf{%
(r})}{\partial \xi _{a}}\times \frac{\partial \mathbf{m}_{\text{st}}%
\mathbf{(r})}{\partial \xi _{b}}\right] .  \label{gyrotropic}
\end{equation}%
We show below that $\boldsymbol{\mathfrak{F}}$ and $\tilde{\Gamma}$\ can be
expressed in terms of the scattering matrix. For our subsequent discussions
it is necessary to include a fluctuating force $\boldsymbol{\mathfrak{f}}(t)$
(with $\left \langle \boldsymbol{\mathfrak{f}}(t)\right \rangle =0),$ which
has not been considered in Refs.~\onlinecite{Tretiakov:prl08,Clarke:prb08}.
From Eq. (\ref{LLG_FDT}) if follows the time correlation of $\boldsymbol{%
\mathfrak{f}}$ is white and obeys the fluctuation-dissipation theorem: 
\begin{equation}
\left \langle \mathfrak{f}_{a}(t)\mathfrak{f}_{b}(t^{\prime })\right \rangle
=2k_{B}T\tilde{\Gamma}_{ab}\delta (t-t^{\prime }).  \label{forces_FDT}
\end{equation}%
In the following we illustrate the collective coordinate description of
magnetization textures for the macrospin model and the Walker model for a
transverse domain wall. The treatment is easily extended to other rigid
textures such as magnetic vortices.

\subsection{Macrospin excitations}

When high magnetic fields are applied or when the system dimensions are
small the exchange stiffness dominates. In both limits the magnetization
direction and its low energy excitations lie on the unit sphere and its
magnetization dynamics is described by the polar angles $\theta (t)$ and $%
\varphi (t)$:%
\begin{equation}
\mathbf{m}=\left( \sin \theta \cos \varphi ,\sin \theta \sin \varphi ,\cos
\theta \right) .  \label{m_macrospin}
\end{equation}%
The diagonal components of the gyrotropic tensor vanish by (anti)symmetry $%
\tilde{\eta}_{\theta \theta }=0$, $\tilde{\eta}_{\varphi \varphi }=0.$ Its
off-diagonal components are 
\begin{equation}
\eta _{\theta \varphi }=\frac{M_{s}V}{\gamma }\sin \theta =-\eta _{\varphi
\theta .}
\end{equation}%
$V$ is the particle volume and $M_{s}V$ the total magnetic moment. We now
have two coupled equations of motion 
\begin{align}
\frac{M_{s}V}{\gamma }\dot{\varphi}\sin \theta -\frac{\partial \Omega }{%
\partial \theta }-\left( \tilde{\Gamma}_{\theta \theta }\dot{\theta}+\tilde{%
\Gamma}_{\theta \varphi }\dot{\varphi}\right) & =0, \\
-\frac{M_{s}V}{\gamma }\dot{\theta}\sin \theta -\frac{\partial \Omega }{%
\partial \varphi }-\left( \tilde{\Gamma}_{\varphi \theta }\dot{\theta}+%
\tilde{\Gamma}_{\varphi \varphi }\dot{\varphi}\right) & =0.  \notag
\end{align}%
The thermodynamic potential $\Omega $ determines the ballistic trajectories
of the magnetization. The Gilbert damping tensor $\tilde{\Gamma}_{ab}$ will
be computed below, but when isotropic and local, 
\begin{equation}
\tilde{\Gamma}=\tilde{1}\delta (\mathbf{r}-\mathbf{r}^{\prime })M_{s}\alpha
/\gamma ,  \label{Gammais}
\end{equation}%
where $\tilde{1}$ is a unit matrix in the Cartesian basis and $\alpha $ is
the dimensionless Gilbert constant, $\Gamma _{\theta \theta }=M_{s}V\alpha
/\gamma $, $\Gamma _{\theta \varphi }=0=\Gamma _{\varphi \theta }$, and $%
\Gamma _{\varphi \varphi }=\sin ^{2}\theta M_{s}V\alpha /\gamma $.

\subsection{Domain Wall Motion}

We focus on a one-dimensional model, in which the magnetization gradient,
magnetic easy axis, and external magnetic field point along the wire ($z$)
axis. The magnetic energy of such a wire with transverse cross section $S$
can be written as\cite{Goussev:prl10} 
\begin{equation}
\Omega =M_{s}S\int dz\phi (z),
\end{equation}%
in terms of the one-dimensional energy density 
\begin{equation}
\phi =\frac{A}{2}\left \vert \frac{\partial \mathbf{m}}{\partial z}\right
\vert ^{2}-H_{a}m_{z}+\frac{K_{1}}{2}\left( 1-m_{z}^{2}\right) +\frac{K_{2}}{%
2}m_{x}^{2},  \label{DW_energy}
\end{equation}%
where $H_{a}$ is the applied field and $A$ is the exchange stiffness. Here
the easy-axis anisotropy is parametrized by an anisotropy constant $K_{1}$.
In the case of a thin film wire, there is also a smaller anisotropy energy
associated with the magnetization transverse to the wire governed by $K_{2}$%
. In a cylindrical wire from a material without crystal anisotropy (such as
permalloy) $K_{2}=0$.

When the shape of such a domain wall is preserved in the dynamics, three
collective coordinates characterize the magnetization texture: the domain
wall position\ $\xi _{1}(t)=r_{\text{w}}(t)$, the polar angle $\xi
_{2}(t)=\varphi _{\text{w}}(t)$, and the domain wall width $\lambda _{%
\text{w}}(t)$. We consider a head-to-head transverse domain wall (a
tail-to-tail wall can be treated analogously). $\mathbf{m}(z)=\left( \sin
\theta _{\text{w}}\cos \varphi _{\text{w}},\sin \theta _{\text{w}}\sin
\varphi _{\text{w}},\cos \theta _{\text{w}}\right) $, where%
\begin{equation}
\cos \theta _{\text{w}}=\tanh \frac{r_{\text{w}}-z}{\lambda _{\text{w}}%
}
\end{equation}%
and 
\begin{equation}
\csc \theta _{\text{w}}=\cosh \frac{r_{\text{w}}-z}{\lambda _{\text{w}}%
}
\end{equation}%
minimizes the energy (\ref{DW_energy}) under the constraint that the
magnetization to the far left and right points towards the domain wall. The
off-diagonal elements are then $\tilde{\eta}_{rl}=0=\tilde{\eta}_{lr}$ and $%
\tilde{\eta}_{r\varphi }=-2M_{s}/\gamma =-\tilde{\eta}_{\varphi r}.$ The
energy (\ref{DW_energy}) reduces to 
\begin{equation}
\Omega =M_{s}S\left[ A/\lambda _{\text{w}}-2H_{a}r+K_{1}\lambda _{\text{w%
}}+K_{2}\lambda _{\text{w}}\cos ^{2}\varphi _{\text{w}}\right] .
\end{equation}%
Disregarding fluctuations, the equation of motion Eq.\ (\ref%
{equation_collective}) can be expanded as: 
\begin{align}
2\dot{r}_{\text{w}}+\alpha _{\varphi \varphi }\dot{\varphi}+\alpha
_{\varphi r}\dot{r}_{\text{w}}+\alpha _{\varphi \lambda }\dot{\lambda}_{%
\text{w}}& =\gamma K_{2}\lambda _{\text{w}}\sin 2\varphi _{\text{w}},
\\
-2\dot{\varphi}+\alpha _{rr}\dot{r}_{\text{w}}+\alpha _{r\varphi }\dot{%
\varphi}+\alpha _{r\lambda }\dot{\lambda}_{\text{w}}& =2\gamma H_{a}, \\
A/\lambda _{\text{w}}^{2}+\alpha _{\lambda r}\dot{r}_{\text{w}}+\alpha
_{\lambda \varphi }\dot{\varphi}+\alpha _{\lambda \lambda }\dot{\lambda}_{%
\text{w}}& =K_{1}+K_{2}\cos ^{2}\varphi _{\text{w}},
\end{align}%
where $\alpha _{ab}=\gamma \Gamma _{ab}/M_{s}S$.

When the Gilbert damping tensor is isotropic and local in the basis of the
Cartesian coordinates, $\tilde{\Gamma}=\tilde{1}\delta (\mathbf{r}-\mathbf{r}%
^{\prime })M_{s}\alpha /\gamma $%
\begin{equation}
\alpha _{rr}=\frac{2\alpha }{\lambda _{\text{w}}};\; \alpha _{\varphi
\varphi }=2\alpha \lambda _{\text{w}};\; \alpha _{\lambda \lambda }=\frac{%
\pi ^{2}\alpha }{6\lambda _{\text{w}}}.
\end{equation}%
whereas all off-diagonal elements vanish.

Most experiments are carried out on thin film ferromagnetic wires for which $%
K_{2}$ is finite. Dissipation is especially simple below the Walker
threshold, the regime in which the wall moves with a constant drift
velocity, $\dot{\varphi}_{\text{w}}=0$ and\cite{Schryer:jap74}%
\begin{equation}
\dot{r}_{\text{w}}=-2\gamma H_{a}/\alpha _{rr}.
\end{equation}%
The Gilbert damping coefficient $\alpha _{rr}$ can be obtained directly from
the scattering matrix by the parametric dependence of the scattering matrix
on the center coordinate position $r_{\text{w}}$. When the Gilbert damping
tensor is isotropic and local, we find $\dot{r}_{\text{w}}=\lambda _{%
\text{w}}\gamma H_{a}/\alpha $. The domain wall width $\lambda _{\text{w}%
}=\sqrt{A/(K_{1}+K_{2}\cos ^{2}\varphi _{\text{w}})}$ and the out-of-plane
angle $\varphi _{\text{w}}=%
{\frac12}%
\arcsin 2\gamma H_{a}/\alpha K_{2}$. At the Walker-breakdown field $\left(
H_{a}\right) _{\text{WB}}=\alpha K_{2}/\left( 2\gamma \right) $ the
sliding domain wall becomes unstable.

In a cylindrical wire without anisotropy, $K_{2}=0$, $\varphi _{\text{w}}\ 
$is time-dependent and satisfies 
\begin{equation}
\dot{\varphi}_{\text{w}}=-\frac{\left( 2+\alpha _{\varphi r}\right) }{%
\alpha _{\varphi \varphi }}\dot{r}_{\text{w}}
\end{equation}%
while 
\begin{equation}
\dot{r}_{\text{w}}=\frac{2\gamma H_{a}}{2\left( \frac{2+\alpha _{\varphi r}%
}{\alpha _{\varphi \varphi }}\right) +\alpha _{rr}}.
\end{equation}%
For isotropic and local Gilbert damping coefficients,\cite{Goussev:prl10} 
\begin{equation}
\frac{\dot{r}_{\text{w}}}{\lambda _{\text{w}}}=\frac{\alpha \gamma H_{a}%
}{1+\alpha ^{2}}.
\end{equation}%
In the next section, we formulate how the Gilbert scattering tensor can be
computed from time-dependent scattering theory.

\section{Scattering theory of mesoscopic magnetization dynamics}

\label{Scattering}

Scattering theory of transport phenomena\cite{Nazarov:09} has proven its
worth in the context of magnetoelectronics. It has been used advantageously
to evaluate the nonlocal exchange interactions multilayers or spin valves,%
\cite{Bruno:prb95} the giant magnetoresistance,\cite{Bauer:prl92}
spin-transfer torque,\cite{Brataas:prl00} and spin pumping.\cite%
{Tserkovnyak:prl02} We first review the scattering theory of equilibrium
magnetic properties and anisotropy fields and then will turn to
non-equilibrium transport.

\subsection{Conservative forces}

Considering only the electronic degrees of freedom in our model, the
thermodynamic (grand) potential is defined as 
\begin{equation}
\Omega =-k_{B}T\ln \text{Tr}e^{-(\hat{H}_{\text{ALDA}}-\mu \hat{N})},
\end{equation}%
while $\mu $ is the chemical potential, and $\hat{N}$ is the number
operator. The conservative force 
\begin{equation}
\boldsymbol{\mathfrak{F}}=-\frac{\partial \Omega }{\partial \boldsymbol{\xi }%
}.  \label{Force_general}
\end{equation}%
can be computed for an open systems by defining a scattering region that is
connected by ideal leads to reservoirs at common equilibrium. For a
two-terminal device, the flow of charge, spin, and energy between the
reservoirs can then be described in terms of the $S$-matrix:%
\begin{equation}
S=\left( 
\begin{array}{cc}
\mathfrak{r} & \mathfrak{t}^{\prime } \\ 
\mathfrak{t} & \mathfrak{r}^{\prime }%
\end{array}%
\right) ,
\end{equation}%
where $\mathfrak{r}$ is the matrix of probability amplitudes of states
impinging from and reflected into the left reservoir, while $\mathfrak{t}$
denotes the probability amplitudes of states incoming from the left and
transmitted to the right. Similarly, $\mathfrak{r}^{\prime }$ and $\mathfrak{%
t}^{\prime }$ describes the probability amplitudes for states that originate
from the right reservoir. $\mathfrak{r}$, $\mathfrak{r}^{\prime }$, $%
\mathfrak{t}$, and $\mathfrak{t}^{\prime }$ are matrices in the space
spanned by eigenstates in the leads. We are interested in the free magnetic
energy modulation by the magnetic configuration that allows evaluation of
the forces Eq.\ (\ref{Force_general}). The free energy change reads 
\begin{equation}
\Delta \Omega =-k_{B}T\int d\epsilon \Delta n(\epsilon )\ln \left[
1+e^{(\epsilon -\mu )/k_{B}T}\right] ,
\end{equation}%
where $\Delta n(\epsilon )d\epsilon $ is the change in the number of states
at energy $\epsilon $ and interval $d\epsilon $, which can be expressed in
terms of the scattering matrix\cite{Avishai:prb85}%
\begin{equation}
\Delta n(\epsilon )=-\frac{1}{2\pi i}\frac{\partial }{\partial \epsilon }%
\text{Tr}\ln S(\epsilon ).
\end{equation}%
Carrying out the derivative, we arrive at the force 
\begin{equation}
\boldsymbol{\mathfrak{F}}=-\frac{1}{2\pi i}\int d\epsilon f(\epsilon )%
\text{Tr}\left( S^{\dagger }\frac{\partial S}{\partial \boldsymbol{\xi }}%
\right) \,,  \label{force_average}
\end{equation}%
where $f(\epsilon )$ is the Fermi-Dirac distribution function with chemical
potential $\mu $. This established result will be reproduced and generalized
to the description of dissipation and fluctuations below.

\subsection{Gilbert damping as energy pumping}

Here we interpret Gilbert damping as an energy pumping process by equating
the results for energy dissipation from the microscopic adiabatic pumping
formalism with the LLG phenomenology in terms of collective coordinates,
Eq.\ (\ref{energy_dissipation_collective}). The adiabatic energy loss rate
of a scattering region in terms of scattering matrix at zero temperature has
been derived in Refs.~\onlinecite{Avron:prl01,Moskalets:prb02}. In the appendices, we generalize this result to finite temperatures:
\begin{equation}
\dot{E}=\frac{\hbar }{4\pi }\int d\epsilon \left( -\frac{\partial f}{%
\partial \epsilon }\right) \text{Tr}\left[ \frac{\partial S(\epsilon ,t)}{%
\partial t}\frac{\partial S^{\dag }(\epsilon ,t)}{\partial t}\right] .
\label{energypumpingformula}
\end{equation}%
Since we
employ the adiabatic approximation, $S(\epsilon ,t)$ is the energy-dependent
scattering matrix for an instantaneous (\textquotedblleft
frozen\textquotedblright ) scattering potential at time $t$. In a magnetic
system, the time dependence arises from its magnetization dynamics, $%
S(\epsilon ,t)=S[\mathbf{m}(t)](\epsilon )$. In terms of the collective
coordinates $\boldsymbol{\xi }(t)$, $S(\epsilon ,t)=S(\epsilon ,\left \{ 
\mathbf{\xi }(t)\right \} )$ 
\begin{equation}
\frac{\partial S[\mathbf{m}(t)]}{\partial t}\approx \sum \limits_{a}\frac{%
\partial S}{\partial \xi _{a}}\dot{\xi}_{a}\,,
\end{equation}%
where the approximate sign has been discussed in the previous section. We
can now identify the dissipation tensor (\ref{diss_tensor}) in terms of the
scattering matrix%
\begin{equation}
\Gamma _{ab}=\frac{\hbar }{4\pi }\int d\epsilon \left( -\frac{\partial f}{%
\partial \epsilon }\right) \text{Tr}\left[ \frac{\partial S(\epsilon )}{%
\partial \xi _{a}}\frac{\partial S^{\dag }(\epsilon )}{\partial \xi _{b}}%
\right] \,.  \label{diss_tensor_pumping}
\end{equation}%
In the macrospin model the Gilbert damping tensor can then be expressed as%
\begin{equation}
\tilde{\alpha}_{ij}=\frac{\gamma \hbar }{4\pi M_{s}}\int d\epsilon \left( -%
\frac{\partial f}{\partial \epsilon }\right) \text{Tr}\left[ \frac{%
\partial S(\epsilon )}{\partial m_{i}}\frac{\partial S^{\dag }(\epsilon )}{%
\partial m_{j}}\right] ,  \label{Gilbert_macrospin}
\end{equation}%
where $m_{i}$ is a Cartesian component of the magnetization direction..

\subsection{Gilbert damping and fluctuation-dissipation theorem}

At finite temperatures the forces acting on the magnetization contain
thermal fluctuations that are related to the Gilbert dissipation by the
fluctuation-dissipation theorem, Eq.\ (\ref{forces_FDT}). The dissipation
tensor is therefore accessible via the stochastic forces in thermal
equilibrium.

The time dependence of the force operators 
\begin{equation}
\hat{\boldsymbol{\mathfrak{F}}}(t)=-\frac{\partial \hat{H}_{\text{ALDA}}(%
\mathbf{m})}{\partial \boldsymbol{\xi }}
\end{equation}%
is caused by the thermal fluctuations of the magnetization. It is convenient
to rearrange the Hamiltonian $\hat{H}_{\text{ALDA}}$ into an unperturbed part that does
not depend on the magnetization and a scattering potential $\hat{H}_{\text{%
ALDA}}(\mathbf{m})=\hat{H}_{0}+\hat{V}(\mathbf{m})$. In the basis of
scattering wave functions of the leads, the force operator reads 
\begin{widetext}
\begin{equation}
\hat{\boldsymbol{\mathfrak{F}}}=-\int d\epsilon \int d\epsilon ^{\prime }\langle \epsilon
\alpha |\frac{\partial \hat{V}}{\partial \boldsymbol{\xi }}|\epsilon ^{\prime }\beta
\rangle \hat{a}_{\alpha }^{\dag }(\epsilon )\hat{a}_{\beta }(\epsilon
^{\prime })e^{i(\epsilon -\epsilon ^{\prime })t /\hbar },
\label{force_operator}
\end{equation}
\end{widetext}where $\hat{a}_{\beta }$ annihilates an electron incident on
the scattering region, $\beta $ labels the lead (left or right) and quantum
numbers of the wave guide mode, and $|\epsilon ^{\prime }\beta \rangle $ is
an associated scattering eigenstate at energy $\epsilon ^{\prime }$. We take
again the left and right reservoirs to be in thermal equilibrium with the
same chemical potentials, such that the expectation values 
\begin{equation}
\left \langle \hat{a}_{\alpha }^{\dag }(\epsilon )\hat{a}_{\beta }(\epsilon
^{\prime })\right \rangle =\delta _{\alpha \beta }\delta (\epsilon -\epsilon
^{\prime })f(\epsilon ).
\end{equation}%
The relation between the matrix element of the scattering potential and the $%
S$-matrix 
\begin{equation}
\left[ S^{\dagger }(\epsilon )\frac{\partial S(\epsilon )}{\partial 
\boldsymbol{\xi }}\right] _{\alpha \beta }=-2\pi i\langle \epsilon \alpha |%
\frac{\partial \hat{V}}{\partial \boldsymbol{\xi }}|\epsilon \beta \rangle \,
\end{equation}%
follows from the relation derived in Eq. (\ref{changeSmatrix}) below as well
as unitarity of the $S$-matrix, $S^{\dagger }S=1$. Taking these relations
into account, the expectation value of $\hat{\boldsymbol{\mathfrak{F}}}$ is
found to be Eq.~(\ref{force_average}). We now consider the fluctuations in
the force $\hat{\boldsymbol{\mathfrak{f}}}(t)=\hat{\boldsymbol{\mathfrak{F}}}%
(t)-\langle \hat{\boldsymbol{\mathfrak{F}}}(t)\rangle $, which involves
expectation values 
\begin{align}
& \left \langle \hat{a}_{\alpha _{1}}^{\dag }(\epsilon _{1})\hat{a}_{\beta
_{1}}(\epsilon _{1}^{\prime })\hat{a}_{\alpha _{2}}^{\dag }(\epsilon _{2})%
\hat{a}_{\beta _{2}}(\epsilon _{2}^{\prime })\right \rangle  \notag \\
& -\left \langle \hat{a}_{\alpha _{1}}^{\dag }(\epsilon _{1})\hat{a}_{\beta
_{1}}(\epsilon _{1}^{\prime })\right \rangle \left \langle \hat{a}_{\alpha
_{2}}^{\dag }(\epsilon _{2})\hat{a}_{\beta _{2}}(\epsilon _{2}^{\prime
})\right \rangle  \notag \\
& =\delta _{\alpha _{1}\beta _{2}}\delta \left( \epsilon _{1}-\epsilon
_{2}^{\prime }\right) \delta _{\beta _{1}\alpha _{2}}\delta \left( \epsilon
_{1}^{\prime }-\epsilon _{2}\right) f(\epsilon _{1})\left[ 1-f(\epsilon _{2})%
\right] \,,
\end{align}%
where we invoked Wick's theorem. Putting everything together, we finally
find \ 
\begin{equation}
\left \langle \mathfrak{f}_{a}(t)\mathfrak{f}_{b}(t^{\prime })\right \rangle
=2k_{B}T\delta (t-t^{\prime })\Gamma _{ab},
\end{equation}%
where $\Gamma _{ab}$ has been defined in Eq.\ (\ref{diss_tensor_pumping}).
Comparing with Eq. (\ref{forces_FDT}), we conclude that the dissipation
tensor $\Gamma _{ab}$ governing the fluctuations is identical to the one
obtained from the energy pumping, Eq. (\ref{diss_tensor_pumping}), thereby
confirming the fluctuation-dissipation theorem.

\section{Kubo Formula}

\label{Kubo}

The quality factor of the magnetization dynamics of most ferromagnets is
high ($\alpha \lesssim 0.01$). Damping can therefore often be treated as a
small perturbation. In the present Section we demonstrate that the damping
obtained from linear response (Kubo) theory agrees\cite{Mahan} with that of
the scattering theory of magnetization dissipation in this limit. At
sufficiently low temperatures or strong elastic disorder scattering the
coupling to phonons may be disregarded and is not discussed here.

The energy dissipation can be written as 
\begin{equation}
\dot{E}=\left \langle \frac{d\hat{H}}{dt}\right \rangle ,
\end{equation}%
where $\left \langle {}\right \rangle $ denotes the expectation value for the
non-equilibrium state. We are interested in the adiabatic response of the
system to a time-dependent perturbation. In the adiabatic (slow) regime, we
can \textit{at any time} expand the Hamiltonian around a static
configuration at the reference time $t=0$,  
\begin{equation}
\hat{H}=\hat{H}_{\text{st}}+\sum_{a}\delta \xi _{a}(t)\left( \frac{%
\partial \hat{H}}{\partial \xi _{a}}\right) _{\mathbf{m}(\mathbf{r}%
)\rightarrow \mathbf{m}_{\text{st}}(\mathbf{r})}.
\end{equation}
The static part, $\hat{H}_{\text{st}}$, is the Hamiltonian for a
magnetization for a fixed and arbitrary initial texture $\mathbf{m}_{\text{%
st}}$, as, without loss of generality, described by the collective
coordinates $\xi _{a}$. Since we assume that the variation of the
magnetization in time is small, a linear expansion in terms of the small
deviations of the collective coordinate $\delta \xi _{i}(t)$ is valid for
sufficiently short time intervals. We can then employ the Kubo formalism and
express the energy dissipation as 
\begin{equation}
\dot{E}=\sum_{a}\delta \dot{\xi}_{a}(t)\left( \frac{\partial \hat{H}}{%
\partial \xi _{a}}\right) _{\mathbf{m}(\mathbf{r})\rightarrow \mathbf{m}_{%
\text{st}}(\mathbf{r})}\,,
\end{equation}%
where the expectation value of the out-of-equilibrium conservative force%
\begin{equation}
\left( \frac{\partial \hat{H}}{\partial \xi _{a}}\right) _{\mathbf{m}(%
\mathbf{r})\rightarrow \mathbf{m}_{\text{st}}(\mathbf{r})}\equiv \partial
_{a}\hat{H}
\end{equation}%
consists of an equilibrium contribution and a term linear in the perturbed
magnetization direction: 
\begin{equation}
\left \langle \partial _{a}\hat{H}\right \rangle (t)=\left \langle \partial _{a}%
\hat{H}\right \rangle _{\text{st}}+\sum \limits_{b}\int_{-\infty }^{\infty
}dt^{\prime }\chi _{ab}(t-t^{\prime })\delta \xi _{b}(t^{\prime })\,.
\end{equation}%
Here, we introduced the retarded susceptibility 
\begin{equation}
\chi _{ab}(t-t^{\prime })=-\frac{i}{\hbar }\theta (t-t^{\prime
})\left \langle \left[ \partial _{a}\hat{H}(t),\partial _{b}\hat{H}(t^{\prime
})\right] \right \rangle _{\text{st}}\,,
\end{equation}%
where $\left \langle {}\right \rangle _{\text{st}}$ is the expectation value
for the wave functions of the static configuration. Focussing on slow
modulations we can further simplify the expression by expanding  
\begin{equation}
\delta \xi _{a}(t^{\prime })\approx \delta \xi _{a}(t)+\left( t^{\prime
}-t\right) \delta \dot{\xi}_{a}(t),
\end{equation}%
so that 
\begin{align}
\left \langle \partial _{a}\hat{H}\right \rangle & =\left \langle \partial _{a}%
\hat{H}\right \rangle _{\text{st}}+\int_{-\infty }^{\infty }dt^{\prime
}\chi _{ab}(t-t^{\prime })\delta \xi _{b}(t)+  \notag \\
& \int_{-\infty }^{\infty }dt^{\prime }\chi _{ab}(t-t^{\prime })\left(
t^{\prime }-t\right) \delta \dot{\xi}_{b}(t).  \label{dev_H}
\end{align}%
The first two terms in this expression, $\langle \partial _{a}\hat{H}\rangle
_{\text{st}}+\int_{-\infty }^{\infty }dt^{\prime }\chi _{ab}(t-t^{\prime
})\delta \xi _{b}(t),$ correspond to the energy variation with respect to a
change in the static magnetization. These terms do not contribute to the
dissipation since the magnetic excitations are transverse, $\mathbf{\dot{m}}%
\cdot \mathbf{m}=0$. Only the last term in Eq. (\ref{dev_H}) gives rise to
dissipation. Hence, the energy loss reduces to\cite{Simanek:prb03} 
\begin{equation}
\dot{E}=i\sum_{ij}\delta \dot{\xi}_{a}\delta \dot{\xi}_{b}\left. \frac{%
\partial \chi _{ab}^{S}}{\partial \omega }\right \vert _{\omega =0},
\end{equation}%
where $\chi _{ab}^{S}(\omega )=\int_{-\infty }^{\infty }dt\left[ \chi
_{ab}(t)+\chi _{ba}(t)\right] e^{i\omega t}/2$. The symmetrized
susceptibility can be expanded as 
\begin{equation}
\chi _{ab}^{S}=\sum \limits_{nm}\frac{\left( f_{n}-f_{m}\right) }{2}\frac{%
\langle n|\partial _{a}\hat{H}|m\rangle \langle m|\partial _{b}\hat{H}%
|n\rangle +(a\leftrightarrow b)}{\hbar \omega +i\eta -(\epsilon
_{n}-\epsilon _{m})},
\end{equation}%
where $|n\rangle $ is an eigenstate of the Hamiltonian $\hat{H}_{\text{st}}
$ with eigenvalue $\epsilon _{n}$, $f_{n}\equiv f(\epsilon _{n})$, $%
f(\epsilon )$ is the Fermi-Dirac distribution function at energy $\epsilon $%
, and $\eta $ is a positive infinitesimal constant. Therefore, 
\begin{widetext}
\begin{equation}
i\left( \frac{\partial \chi _{ab}^{S}}{\partial \omega }\right) _{\omega
=0}=\pi \sum \limits_{nm}\left( -\frac{\partial f_{n}}{\partial \epsilon }%
\right) \langle n|\partial _{a}\hat{H}|m\rangle \langle m|\partial _{b}\hat{H%
}|n\rangle \delta (\epsilon _{n}-\epsilon _{m}),
\end{equation}
and the dissipation tensor 
\begin{equation}
\Gamma _{ab}=\pi \sum \limits_{nm}\left( -\frac{\partial f_{n}}{\partial
\epsilon }\right) \langle n|\partial _{a}\hat{H}|m\rangle \langle m|\partial
_{b}\hat{H}|n\rangle \delta (\epsilon _{n}-\epsilon _{m}).
\label{diss_tensor_Kubo}
\end{equation}%
\end{widetext}We now demonstrate that the dissipation tensor obtained from
the Kubo linear response formula, Eq. (\ref{diss_tensor_Kubo}), is identical
to the expression from scattering theory, Eq.\ (\ref{diss_tensor_pumping}),
following the Fisher and Lee proof of the equivalence of linear response and
scattering theory for the conductance.\cite{Fisher:prb81}

The static Hamiltonian $\hat{H}_{\text{st}}(\mathbf{\xi })=\hat{H}_{0}+%
\hat{V}(\mathbf{\xi })$ can be decomposed into a free-electron part $\hat{H}%
_{0}=-\hbar ^{2}\nabla ^{2}/2m$ and a scattering potential $\hat{V}(\mathbf{%
\xi })$. The eigenstates of $\hat{H}_{0}$ are denoted $\left \vert \varphi
_{s,q}(\epsilon )\right \rangle ,$ with eigenenergies $\epsilon $, where $%
s=\pm $ denotes the longitudinal propagation direction along the system
(say, to the left or to the right), and $q$ a transverse quantum number
determined by the lateral confinement. The potential $\hat{V}(\mathbf{\xi })$
scatters the particles between the propagating states forward or backward.
The outgoing ($+$) and incoming ($-$) scattering eigenstates of the static
Hamiltonian $\hat{H}_{\text{st}}$ are written as $\left \vert \psi
_{s,q}^{(\pm )}(\epsilon )\right \rangle $, which form another complete basis
with orthogonality relations $\left \langle \psi _{s,q}^{(\pm )}(\epsilon
)\right. \left \vert \psi _{s^{\prime },q^{\prime }}^{(\pm )}(\epsilon
^{\prime })\right \rangle =\delta _{s,s^{\prime }}\delta _{q,q^{\prime
}}\delta (\epsilon -\epsilon ^{\prime })$. \cite{MelloKumar:book04} These
wave functions can be expressed as $\left \vert \psi _{s,q}^{(\pm )}(\epsilon
)\right \rangle =[1+\hat{G}_{\text{st}}^{(\pm )}\hat{V}]\left \vert \varphi
_{s,q}\right \rangle $, where the retarded ($+$) and advanced ($-$) Green's
functions read $\hat{G}_{\text{st}}^{(\pm )}(\epsilon )=(\epsilon \pm
i\eta -\hat{H}_{\text{st}})^{-1}$. By expanding $\Gamma _{ab}$ in the
basis of outgoing wave functions, $|\psi _{s,q}^{(+)}\rangle $ , the energy
dissipation (\ref{diss_tensor_Kubo}) becomes 
\begin{widetext}
\begin{equation}
\Gamma _{ab}=\pi  \sum_{sq,s^{\prime
}q^{\prime }}\int d\epsilon \left( -\frac{\partial f_{s,q}}{\partial \epsilon }\right)  \left \langle \psi _{s,q}^{(+)}\right|\partial _{a}\hat{H}\left|\psi
_{s^{\prime },q^{\prime }}^{(+)}\right \rangle \left \langle \psi _{s^{\prime
},q^{\prime }}^{(+)}\right|\partial _{b}\hat{H}\left|\psi _{s,q}^{(+)}\right \rangle ,
\label{diss_tensor_Kubo_scattering}
\end{equation}%
where wave functions should be evaluated at the energy $\epsilon $.

Let us now compare this result, Eq.~(\ref{diss_tensor_Kubo_scattering}), to the
direct scattering matrix expression for the energy dissipation, Eq.~(\ref%
{diss_tensor_pumping}). The $S$-matrix operator can be written in terms of the $T$-matrix as $\hat{S}(\epsilon ;\mathbf{\xi })=1-2\pi i%
\hat{T}(\epsilon ;\mathbf{\xi })$, where the $T$-matrix is defined recursively by $%
\hat{T}=\hat{V}[ 1+\hat{G}_{\rm st}^{(+)}\hat{T}] $.
We then find 
\begin{equation*}
\frac{\partial \hat{T}}{\partial \xi _{a}}=\left[ 1+\hat{V}\hat{G}_{\text{st%
}}^{(+)}\right] \partial _{a}\hat{H}\left[ 1+\hat{G}_{\text{st}}^{(+)}\hat{V%
}\right] \,.
\end{equation*}%
The change in the scattering matrix appearing in Eq.~(\ref{diss_tensor_pumping}) is then 
\begin{equation}
\frac{\partial S_{s^{\prime }q^{\prime },sq}}{\partial \xi _{a}} =-2\pi i\left \langle
\varphi _{s,q}\right|\left[ 1+\hat{V}\hat{G}_{\text{st}}^{(+)}\right] \partial
_{a}\hat{H}\left[ 1+\hat{G}_{\text{st}}^{(+)}\hat{V}\right] |\varphi
_{s^{\prime },q^{\prime }}\rangle   =-2\pi i\left \langle \psi _{s^{\prime },q^{\prime }}^{(-)}\right|\partial _{a}\hat{H}%
\left|\psi _{s^{\prime },q^{\prime }}^{(+)}\right \rangle .\label{changeSmatrix}
\end{equation}%
\end{widetext}Since 
\begin{equation}
\left \langle \psi _{s,q}^{(-)}(\epsilon )\right \vert
=\sum \limits_{s^{\prime }q^{\prime }}S_{sq,s^{\prime }q^{\prime
}}\left \langle \psi _{s^{\prime }q^{\prime }}^{(+)}(\epsilon )\right \vert
\end{equation}%
and $SS^{\dag }=1$, we can write the linear response result, Eq.~(\ref%
{diss_tensor_Kubo_scattering}), as energy pumping (\ref{diss_tensor_pumping}%
). This completes our proof of the equivalence between adiabatic energy
pumping in terms of the $S$-matrix and the Kubo linear response theory.

\section{Conclusions}

\label{Conclusions}

We have shown that most aspects of magnetization dynamics in ferromagnets
can be understood in terms of the boundary conditions to normal metal
contacts, \textit{i.e.} a scattering matrix. By using the established
numerical methods to compute electron transport based on scattering theory,
this opens the way to compute dissipation in ferromagnets from
first-principles. In particular, our formalism should work well for systems
with strong elastic scattering due to a high density of large impurity
potentials or in disordered alloys, including Ni$_{1-x}$Fe$_{x}$ ($x=0.2$
represents the technologically important \textquotedblleft
permalloy\textquotedblright ).

The dimensionless Gilbert damping tensors (\ref{Gilbert_macrospin}) for
macrospin excitations, which can be measured directly in terms of the
broadening of the ferromagnetic resonance, have been evaluated for Ni$_{1-x}$%
Fe$_{x}$ alloys by \textit{ab initio} methods.\cite{Starikov:10} Permalloy
is substitutionally disordered and damping is dominated by the spin-orbit
interaction in combination with disorder scattering. Without adjustable
parameters good agreement has been obtained with the available low
temperature experimental data, which is a strong indication of the practical
value of our approach.

In clean samples and at high temperatures, the electron-phonon scattering
importantly affects damping. Phonons are not explicitly included here, but
the scattering theory of Gilbert damping can still be used for a frozen
configuration of thermally displaced atoms, neglecting the inelastic aspect
of scattering.\cite{Liu:10} 

While the energy pumping by scattering theory
has been applied to described magnetization damping,\cite{Brataas:prl08} it
can be used to compute other dissipation phenomena. This has recently been
demonstrated for the case of current-induced mechanical forces and damping,%
\cite{Bode} with a formalism analogous to that for current-induced
magnetization torques.\cite{Hals:prl09,Hals:epl10}

\begin{acknowledgments}
We would like to thank Kjetil Hals, Paul J. Kelly, Yi\ Liu, Hans Joakim
Skadsem, Anton Starikov, and Zhe Yuan for stimulating discussions. This work
was supported by the EC Contract ICT-257159 \textquotedblleft
MACALO,\textquotedblright \ the NSF under Grant No.~DMR-0840965, DARPA, FOM,
DFG, and by the Project of Knowledge Innovation Program (PKIP) of Chinese
Academy of Sciences, Grant No. KJCX2.YW.W10
\end{acknowledgments}

\appendix

\section{Adiabatic Pumping}

Adiabatic pumping is the current response to a time-dependent scattering
potential to first order in the time-variation or \textquotedblleft
pumping\textquotedblright \ frequency when all reservoirs are at the same
electro-chemical potential.\cite{Buttiker:zphysb94} A compact formulation of
the pumping charge current in terms of the instantaneous scattering matrix
was derived in Ref.\  \onlinecite{Brouwer:prb98}. In the same spirit, the
energy current pumped out of the scattering region has been formulated (at
zero temperature) in Ref. \onlinecite{Moskalets:prb02}. Some time ago, we
extended the charge pumping concept to include the spin degree of freedom
and ascertained its importance in magnetoelectronic circuits.\cite%
{Tserkovnyak:prl02} More recently, we demonstrated that the energy emitted
by a ferromagnet with time-dependent magnetizations into adjacent conductors
is not only caused by interface spin pumping, but also reflects the energy
loss by spin-flip processes inside the ferromagnet\cite{Brataas:prl08} and
therefore Gilbert damping. Here we derive the energy pumping expressions at
finite temperatures, thereby generalizing the zero temperature results
derived in Ref.\  \onlinecite{Moskalets:prb02} and used in Ref.\ %
\onlinecite{Brataas:prl08}. Our results differ from an earlier extension to
finite temperature derived in Ref. \onlinecite{Wang:prb02} and we point out
the origin of the discrepancies. The magnetization dynamics must satisfy the
fluctuation-dissipation theorem, which is indeed the case in our formulation.

We proceed by deriving the charge, spin, and energy currents in terms of the
time dependence of the scattering matrix of a two-terminal device. The
transport direction is $x$ and the transverse coordinates are $\boldsymbol{%
\varrho }=(y,z)$. An arbitrary single-particle Hamiltonian can be decomposed
as 
\begin{equation}
H(\mathbf{r})=-\frac{\hbar ^{2}}{2m}\frac{\partial ^{2}}{\partial x^{2}}%
+H_{\perp }(x,\boldsymbol{\varrho }),
\end{equation}%
where the transverse part is 
\begin{equation}
H_{\perp }(x,\boldsymbol{\varrho })=-\frac{\hbar ^{2}}{2m}\frac{\partial ^{2}%
}{\partial \boldsymbol{\varrho }^{2}}+V(x,\boldsymbol{\varrho })\,.
\end{equation}%
$V(\boldsymbol{\varrho })$ is an elastic scattering potential in $2\times 2$
Pauli spin space that includes\ the lattice, impurity, and self-consistent
exchange-correlation potentials, including spin-orbit interaction and
magnetic disorder. The scattering region is attached to perfect non-magnetic
electron wave guides (left $\alpha =L$ and right $\alpha =R$) with constant
potential and without spin-orbit interaction. In lead $\alpha $, the
transverse part of the $2\times 2$ spinor wave function $\varphi _{\alpha
}^{(n)}(x,\boldsymbol{\varrho })$ and its corresponding transverse energy $%
\epsilon _{\alpha }^{(n)}$ obey the Schr\"{o}dinger equation 
\begin{equation}
H_{\perp }(\boldsymbol{\varrho })\varphi _{\alpha }^{(n)}(\boldsymbol{%
\varrho })=\epsilon _{\alpha }^{(n)}\varphi _{\alpha }^{(n)}(\boldsymbol{%
\varrho }),
\end{equation}%
where $n$ is the spin and orbit quantum number. These transverse wave guide
modes form the basis for the expansion of the time-dependent scattering
states in lead $\alpha =L,R$: 
\begin{equation}
\hat{\Psi}_{\alpha }=\int_{0}^{\infty }\frac{dk}{\sqrt{2\pi }}\sum_{n\sigma
}\varphi _{\alpha }^{(n)}(\boldsymbol{\varrho })e^{i\sigma kx}e^{-i\epsilon
_{\alpha }^{(nk)}t/\hbar }\hat{c}_{\alpha }^{(nk\sigma )},
\label{psi_waveguide}
\end{equation}%
where $\hat{c}_{\alpha }^{(nk\sigma )}$ annihilates an electron in mode $n$
incident ($\sigma =+$) or outgoing ($\sigma =-$) in lead $\alpha $. The
field operators satisfy the anticommutation relation 
\begin{equation*}
\left \{ \hat{c}_{\alpha }^{(nk\sigma )},\hat{c}_{\beta }^{\dag (n^{\prime
}k^{\prime }\sigma ^{\prime })}\right \} =\delta _{\alpha \beta }\delta
_{nn^{\prime }}\delta _{\sigma \sigma ^{\prime }}\delta (k-k^{\prime }).
\end{equation*}%
The total energy is $\epsilon _{\alpha }^{(nk)}=\hbar ^{2}k^{2}/2m+\epsilon
_{\alpha }^{(n)}$. In the leads the particle, spins, and energy currents in
the transport direction are \textit{\ } 
\begin{subequations}
\label{currents}
\begin{align}
\hat{I}^{(p)}& =\frac{\hbar }{2mi}\int d\boldsymbol{\varrho }\text{Tr}%
_{s}\left( \hat{\Psi}^{\dag }\frac{\partial \hat{\Psi}}{\partial x}-\frac{%
\partial \hat{\Psi}^{\dag }}{\partial x}\hat{\Psi}\right) ,
\label{energycurrent} \\
\boldsymbol{\hat{I}}^{(s)}& =\frac{\hbar }{2mi}\int d\boldsymbol{\varrho }%
\text{Tr}_{s}\left( \hat{\Psi}^{\dag }\boldsymbol{\sigma }\frac{\partial 
\hat{\Psi}}{\partial x}-\frac{\partial \hat{\Psi}^{\dag }}{\partial x}%
\boldsymbol{\sigma }\hat{\Psi}\right) , \\
\hat{I}^{(e)}& =\frac{\hbar }{4mi}\int d\boldsymbol{\varrho }\text{Tr}%
_{s}\left( \hat{\Psi}^{\dag }H\frac{\partial \hat{\Psi}}{\partial x}-\frac{%
\partial \hat{\Psi}^{\dag }}{\partial x}H\hat{\Psi}\right) +\text{H.c.},
\end{align}%
where we suppressed the time $t$ and lead index $\alpha $, $\boldsymbol{%
\sigma }=(\sigma _{x},\sigma _{y},\sigma _{z})$ is a vector of Pauli
matrices, and $\text{Tr}_{s}$ denotes the trace in spin space. Note that
the spin current $\boldsymbol{I}_{s}$ flows in the $x$-direction with
polarization vector $\boldsymbol{I}_{s}/I_{s}$. To avoid dependence on an
arbitrary global potential shift, it is convenient to work with heat $\hat{I}%
^{(q)}$ rather than energy currents $\hat{I}^{(\epsilon )}:$ 
\end{subequations}
\begin{equation}
\hat{I}^{(q)}(t)=\hat{I}^{(\epsilon )}(t)-\mu \hat{I}^{(p)}(t)\,,
\end{equation}%
where $\mu $ is the chemical potential. Inserting the waveguide
representation (\ref{psi_waveguide}) into (\ref{currents}), the particle
current reads\cite{Buttiker:prb92} 
\begin{align}
\hat{I}_{\alpha }^{(p)}& =\frac{\hbar }{4\pi m}\int_{0}^{\infty
}dkdk^{\prime }\sum_{n\sigma \sigma ^{\prime }}\left( \sigma k+\sigma
^{\prime }k^{\prime }\right) \times  \notag \\
& e^{i\left( \sigma k-\sigma ^{\prime }k^{\prime }\right) x}e^{-i\left[
\epsilon _{\alpha }^{(nk)}-\epsilon _{\alpha }^{(nk^{\prime })}\right]
t/\hbar }\hat{c}_{\alpha }^{\dag (nk^{\prime }\sigma ^{\prime })}\hat{c}%
_{\alpha }^{(nk\sigma )}.
\end{align}%
We are interested in the low-frequency limit of the Fourier transforms $%
I_{\alpha }^{(x)}(\omega )=\int_{-\infty }^{\infty }dte^{i\omega t}I_{\alpha
}^{(x)}(t)$. Following Ref. \onlinecite{Buttiker:prb92} we assume long
wavelengths such that only the intervals with $k\approx k^{\prime }$ and $%
\sigma =\sigma ^{\prime }$ contribute. In the adiabatic limit $\omega
\rightarrow 0$ this approach is correct to leading order in $\hbar \omega
/\epsilon _{F},$ where $\epsilon _{F}$ is the Fermi energy.\textit{\ }By
introducing the (current-normalized)\ operator 
\begin{equation}
\hat{c}_{\alpha }^{(n\sigma )}(\epsilon _{\alpha }^{(nk)})=\frac{1}{\sqrt{%
\frac{d\epsilon _{\alpha }^{(nk\sigma )}}{dk}}}\hat{c}_{\alpha }^{(nk\sigma
)},
\end{equation}%
which obey the anticommutation relations 
\begin{widetext}
\begin{equation}
\left \{ \hat{c}_{\alpha }^{(n\sigma )}(\epsilon _{\alpha }),\hat{%
c}_{\beta }^{\dag (n^{\prime }\sigma ^{\prime })}(\epsilon _{\beta
})\right \}=\delta _{\alpha \beta }\delta
_{nn^{\prime }}\delta _{\sigma \sigma ^{\prime }}\delta (\epsilon _{\alpha
}-\epsilon _{\beta }) \label{com}.
\end{equation}%
The charge current can be written as 
\begin{equation}
\hat{I}_{\alpha }^{(c)}(t) =\frac{1}{2\pi \hbar }\int_{\epsilon _{\alpha
}^{(n)}}^{\infty }d\epsilon d\epsilon ^{\prime }\sum_{n\sigma }\sigma e^{-i( \epsilon-\epsilon')
t/\hbar}\hat{c}_{\alpha }^{\dag(n\sigma )}(\epsilon ^{\prime })\hat{c}%
_{\alpha }^{(n\sigma )}(\epsilon ) \label{integ}.
\end{equation}
\end{widetext}We operate in the linear response regime in which applied
voltages and temperature differences as well as the externally induced
dynamics disturb the system only weakly. Transport is then governed by
states close to the Fermi energy. We may therefore extend the limits of the
energy integration in Eq. (\ref{integ}) from \ $(\epsilon _{\alpha
}^{(n)},\infty )$ to $(-\infty $ to $\infty )$. We relabel the annihilation
operators so that $\hat{a}_{\alpha }^{(nk)}=\hat{c}_{\alpha +}^{(nk)}$
denotes particles incident on the scattering region from lead $\alpha $ and $%
\hat{b}_{\alpha }^{(nk)}=\hat{c}_{\alpha -}^{(nk)}$ denotes particles
leaving the scattering region by lead $\alpha $. Using the Fourier
transforms 
\begin{align}
\hat{c}_{\alpha }^{(n\sigma )}(\epsilon )& =\int_{-\infty }^{\infty }dt\hat{c%
}_{\alpha }^{(n\sigma )}(t)e^{i\epsilon t/\hbar },  \label{cFourier1} \\
\hat{c}_{\alpha }^{(n\sigma )}(t)& =\frac{1}{2\pi \hbar }\int_{-\infty
}^{\infty }d\epsilon \hat{c}_{\alpha }^{(n\sigma )}(\epsilon )e^{-i\epsilon
t/\hbar },  \label{cFourier2}
\end{align}%
we obtain in the low-frequency limit\cite{Buttiker:prb92} 
\begin{equation}
\hat{I}_{\alpha }^{(p)}(t)=2\pi \hbar \left[ \hat{a}_{\alpha }^{\dag }(t)%
\hat{a}_{\alpha }(t)-\hat{b}_{\alpha }^{\dag }(t)\hat{b}_{\alpha }(t)\right]
,  \label{Iclowfreq}
\end{equation}%
where $\hat{b}_{\alpha }$ is a column vector of the creation operators for
all wave-guide modes $\{ \hat{b}_{\alpha }^{(n)}\}$. Analogous calculations
lead to the spin current%
\begin{equation}
\boldsymbol{\hat{I}}_{\alpha }^{(s)}=2\pi \hbar \left( \hat{a}_{\alpha
}^{\dag }\boldsymbol{\sigma }\hat{a}_{\alpha }-\hat{b}_{\alpha }^{\dag }%
\boldsymbol{\sigma }\hat{b}_{\alpha }\right)  \label{Islowfreq}
\end{equation}%
and the energy current 
\begin{equation}
\hat{I}_{\alpha }^{(e)}=i\pi \hbar ^{2}\left( \hat{a}_{\alpha }^{\dag }\frac{%
\partial \hat{a}_{\alpha }}{\partial t}-\hat{b}_{\alpha }^{\dag }\frac{%
\partial \hat{b}_{\alpha }}{\partial t}\right) +\text{H.c.}.
\label{IElowfreq}
\end{equation}%
Next, we express the outgoing operators $\hat{b}(t)$ in terms of the
incoming operators $\hat{a}(t)$ via the time-dependent scattering matrix (in
the space spanned by all waveguide modes, including spin and orbit quantum
number): 
\begin{equation}
\hat{b}_{\alpha }(t)=\sum_{\beta }\int dt^{\prime }S_{\alpha \beta
}(t,t^{\prime })\hat{a}_{\beta }(t^{\prime }).  \label{Smatrix}
\end{equation}%
When the scattering region is stationary, $S_{\alpha \beta }(t,t^{\prime })$
only depends on the relative time difference $t-t^{\prime }$, and its
Fourier transform with respect to the relative time is energy independent, 
\textit{i.e.} transport is elastic and can be computed for each energy
separately. For time-dependent problems, $S_{\alpha \beta }(t,t^{\prime })$
also depends on the total time $t+t^{\prime }$ and there is an inelastic
contribution to transport as well. An electron can originate from a lead
with energy $\epsilon $, pick up energy in the scattering region and end up
in the same or the other lead with different energy $\epsilon ^{\prime }$.

The reservoirs are in equilibrium with controlled local chemical potentials
and temperatures. We insert the $S$-matrix (\ref{Smatrix}) into the
expressions for the currents, Eqs.~(\ref{Iclowfreq}), (\ref{Islowfreq}), (%
\ref{IElowfreq}), and use the expectation value at thermal equilibrium%
\begin{equation}
\left \langle \hat{a}_{\alpha }^{\dag (n)}(t_{2})\hat{a}_{\beta
}^{(m)}(t_{1})\right \rangle _{\text{eq}}=\delta _{nm}\delta _{\alpha \beta
}f_{\alpha }(t_{1}-t_{2})/2\pi \hslash ,
\end{equation}%
where $f_{\beta }(t_{1}-t_{2})=(2\pi \hbar )^{-1}\int d\epsilon ^{-i\epsilon
\left( t_{1}-t_{2}\right) /\hbar }f_{\alpha }(\epsilon )$ and $f_{\alpha
}(\epsilon )$ is the Fermi-Dirac distribution of electrons with energy $%
\epsilon $ in the $\alpha $-th reservoir. We then find 
\begin{widetext} 
\begin{align}
2\pi \hbar \left \langle \hat{b}_{\alpha }^{\dag }(t )\hat{b}%
_{\alpha }(t )\right \rangle _{\text{eq}}&=\sum_{\beta }\int dt _{1}dt _{2}S_{\alpha \beta }^{\ast }(t
,t _{2})S_{\alpha \beta }(t ,t _{1})f_{\beta }(t
_{1}-t _{2}), \\
2\pi \hbar \left \langle \hat{b}_{\alpha }^{\dag }(t )\boldsymbol{\sigma }%
\hat{b}_{\alpha }(t )\right \rangle _{\text{eq}}&=\sum_{\beta }\int dt _{1}dt _{2}S_{\alpha \beta }^{\ast }(t
,t _{2})\boldsymbol{\sigma }S_{\alpha \beta }(t ,t _{1})f_{\beta }(t
_{1}-t _{2}), \\
2\pi \hbar \left \langle \hbar \partial _{t }\hat{b}_{\alpha }^{\dag
}(t )\hat{b}_{\alpha }(t )\right \rangle _{\text{eq}}&=\sum_{\beta }\int dt _{1}dt _{2}\left[ \hbar \partial _{t
}S_{\alpha \beta }^{\ast }(t ,t _{2})\right]S_{\alpha
\beta }(t ,t _{1})f_{\beta }(t _{1}-t _{2}).
\end{align}%
Next, we use the Wigner representation (\ref{WignerRep}):
\begin{equation}
S(t ,t')=\frac{1}{2\pi \hbar}\int_{-\infty }^{\infty }d\epsilon S\left(%
\frac{t +t'}{2},\epsilon \right)e^{-i\epsilon (t -t')/\hbar },
\end{equation}%
and by Taylor expanding the Wigner represented S-matrix $S((t +t')/2,\epsilon )$ around $S(t ,\epsilon ),$ $S((t +t')/2,\epsilon )=\sum_{n=0}^{\infty }\partial _{t }^{n}S(t ,\epsilon
)(t'-t )^{n}/(2^{n}n!)$, we find 
\begin{equation}
S(t ,t')=\frac{1}{2\pi \hbar }\int_{-\infty }^{\infty }d\epsilon
e^{-i\epsilon (t -t')/\hbar }e^{i \hbar \partial _{\epsilon }\partial
_{t }/2} S(t ,\epsilon )
\end{equation}%
and 
\begin{equation}
\hbar \partial _{t }S(t ,t')=\frac{1}{2\pi \hbar }\int_{-\infty
}^{\infty }d\epsilon e^{-i\epsilon (t -t')/\hbar }e^{i \hbar \partial
_{\epsilon }\partial _{t }/2}\left( \frac{1}{2} \hbar \partial _{t
}-i\epsilon \right) S(t ,\epsilon ).
\end{equation}%
The factor $1/2$ scaling the term $\hbar \partial_t S(t,\epsilon)$ arises from commuting $\epsilon$ with $e^{i \hbar \partial_\epsilon \partial_t/2}$. The currents can now be evaluated as %
\begin{subequations}
\label{pumpedcurrents}
\begin{align}
I_{\alpha }^{(c)}(t )=&-\frac{1}{2\pi \hbar }\sum_{\beta }\int_{-\infty
}^{\infty }d\epsilon \left[ \left( e^{-i\partial _{\epsilon }\partial _{t
}\hbar /2}S_{\beta \alpha }^{\dag }(\epsilon ,t )\right) \left( e^{i\partial _{\epsilon }\partial _{t }/2\hbar }S_{\alpha
\beta }(\epsilon ,t )\right) f_{\beta }(\epsilon)-f_{\alpha }(\epsilon)\right]
\label{pumpedenergycurrent} \\
\mathbf{I}_{\alpha }^{(s)}(t )=&-\frac{1}{2\pi \hbar }\sum_{\beta
}\int_{-\infty }^{\infty }d\epsilon \left[ \left( e^{-i\partial _{\epsilon
}\partial _{t }\hbar /2}S_{\beta \alpha }^{\dag }(\epsilon ,t )\right) 
\boldsymbol{\sigma }\left( e^{i\partial _{\epsilon }\partial _{t }/2\hbar
}S_{\alpha \beta }(\epsilon ,t )\right) f_{\beta }(\epsilon)\right] \\
I_{\alpha }^{(\epsilon )}(t )=&-\frac{1}{4\pi \hbar }\sum_{\beta
}\int_{-\infty }^{\infty }d\epsilon \left[ \left( e^{-i\partial _{\epsilon
}\partial _{t }/2\hbar }(-i\hbar \partial _{t }/2+\epsilon )S_{\beta
\alpha }^{\dag }(\epsilon ,t )\right)\left( e^{+i\partial
_{\epsilon }\partial _{t }/2\hbar }S_{\alpha \beta }(\epsilon ,t
)\right) f_{\beta }(\epsilon)-\epsilon f_{\alpha }(\epsilon)\right]  \notag \\
& -\frac{1}{4\pi \hbar }\int_{-\infty }^{\infty }d\epsilon \left[ \left(
e^{-i\partial _{\epsilon }\partial _{t }/2\hbar }S_{\beta \alpha }^{\dag
}(\epsilon ,t )\right) \left( e^{i\partial _{\epsilon
}\partial _{t }/2\hbar }(i\hbar \partial _{t }/2+\epsilon )S_{\alpha
\beta }(\epsilon ,t )\right) f_{\beta }(\epsilon)-\epsilon f_{\alpha }(\epsilon)\right],
\end{align}
\end{subequations}
where the adjoint of the S-matrix has elements $S_{\beta \alpha }^{\dag
(n^{\prime },n)}=S_{\alpha \beta }^{\ast (n,n^{\prime })}$ .

We are interested in the average (DC) currents, where simplified expressions
can be found by partial integration over energy and time intervals. We will consider
the total DC currents \textit{out of} the scattering region,
$I^{(\text{out})}=-\sum_{\alpha }I_{\alpha }$, when the electrochemical
potentials in the reservoirs are equal, $f_{\alpha }(\epsilon )=f(\epsilon )$ for all $\alpha$. The averaged pumped
spin and energy currents out of the system in a time interval $\tau$ can be written compactly as
\begin{subequations}
\begin{align}
\label{DCeqpumpedcurrent}
I_{\text{out}}^{(c)}=&\frac{1}{2\pi \hbar \tau}\int_{0}^{\tau}dt
\int d\epsilon \text{Tr}\left \{ \left[ f\left(\epsilon -\frac{%
i\hbar }{2}\frac{\partial }{\partial t }\right)S\right] S^{\dagger }-f(\epsilon
)\right \},  \\
\label{DCeqpumpedspincurrent}
\mathbf{I}_{\text{out}}^{(s)}=&\frac{1}{2\pi \hbar \tau}%
\int_{0}^{\tau}dt \int d\epsilon \text{Tr}\left \{ \boldsymbol{\sigma }\left[
f\left(\epsilon -\frac{i\hbar }{2}\frac{\partial }{\partial t }\right)S\right]
S^{\dagger }\right \}, \\
I_{\text{out}}^{(\epsilon )}=&\frac{1}{2\pi \hbar \tau}%
\int_{0}^{\tau}dt \int d\epsilon \text{Tr}\left \{ \left[
\left(\epsilon -\frac{i\hbar }{2}\frac{\partial }{\partial t }\right)f\left(\epsilon -%
\frac{i\hbar }{2}\frac{\partial }{\partial t }\right)S\right] S^{\dagger
}-\epsilon f(\epsilon )\right \}  \notag \\
& +\frac{1}{2\pi \hbar \tau}\int_{0}^{\tau}dt \int d\epsilon \text{Tr}%
\left \{ \left[ f\left(\epsilon -\frac{i\hbar }{2}\frac{\partial 
}{\partial t }\right)S\right] \left(-i\hbar \frac{\partial S^{\dagger }}{\partial
t }\right)\right \} \,, \label{DCeqpumpedenergycurrent}
\end{align}%
\end{subequations}
\end{widetext}where Tr is the trace over all waveguide modes (spin and
orbital quantum numbers). As shown in Appendix~\ref{consunitarity} the
charge pumped into the reservoirs vanishes for a scattering matrix with a
periodic time dependence when,integrated over one cycle: 
\begin{equation}
I_{\text{out}}^{(p)}=0.
\end{equation}%
This reflects particle conservation; the number of electrons cannot build up
in the scattering region for periodic variations of the system. We can show
that a similar contribution to the energy current, \textit{i.e.} the first
line in Eq.~(\ref{DCeqpumpedenergycurrent}), vanishes, leading to to the
simple expression 
\begin{equation}
I_{\text{out}}^{(e)}=-\frac{i}{2\pi }\int_{0}^{\tau }\frac{dt}{\tau }\int
d\epsilon \text{Tr}\left \{ \left[ f\left( \epsilon -\frac{i\hbar }{2}\frac{%
\partial }{\partial t}\right) S\right] \frac{\partial S^{\dagger }}{\partial
t}\right \} .  \label{IEsimplified}
\end{equation}%
Expanded to lowest order in the pumping frequency the pumped spin current (%
\ref{DCeqpumpedspincurrent}) becomes 
\begin{equation}
\mathbf{I}_{\text{out}}^{(s)}=\frac{1}{2\pi \hbar }\int_{0}^{\tau }\frac{dt%
}{\tau }\int d\epsilon \text{Tr}\left \{ \left( SS^{\dag }f-\frac{i\hbar }{2%
}\frac{\partial S}{\partial t}S^{\dag }\partial _{\epsilon }f\right) 
\boldsymbol{\sigma }\right \}  \label{Is_quadratic}
\end{equation}%
This formula is not the most convenient form to compute the current to
specified order. $SS^{\dag }$ also contains contributions that are linear
and quadratic in the precession frequency since $S(t,\epsilon )$ is the $S$%
-matrix for a time-dependent problem. Instead, we would like to express the
current in terms of the \textit{frozen} scattering matrix $S_{\text{fr}%
}(t,\epsilon )$. The latter is computed for an instantaneous, static
electronic potential. In our case this is determined by a magnetization
configuration that depends parametrically on time: $S_{\text{fr}%
}(t,\epsilon )=S[\mathbf{m}(t),\epsilon ]$. Using unitarity of the
time-dependent $S$-matrix (as elaborated in Appendix~\ref{consunitarity}),
expand it to lowest order in the pumping frequency, and insert it into (\ref%
{Is_quadratic}) leads to\cite{Brouwer:prb98} 
\begin{equation}
\mathbf{I}_{\text{out}}^{(s)}=\frac{i}{2\pi }\sum \limits_{\beta
}\int_{0}^{\tau }\frac{dt}{\tau }\int d\epsilon \left( -\frac{\partial f}{%
\partial \epsilon }\right) \text{Tr}\left \{ \frac{\partial S_{\text{fr}}%
}{\partial t}S_{\text{fr}}^{\dagger }\boldsymbol{\sigma }\right \} .
\end{equation}

We evaluate the energy pumping by expanding (\ref{IEsimplified}) to second
order in the pumping frequency:%
\begin{equation}
I_{\text{out}}^{(e)}=\frac{\hbar }{4\pi }\int_{0}^{\tau }\frac{dt}{\tau }%
\int d\epsilon \text{Tr}\left \{ -ifS\frac{\partial S^{\dagger }}{\partial t%
}-(\partial _{\epsilon }f)\frac{1}{2}\frac{\partial S}{\partial t}\frac{%
\partial S^{\dagger }}{\partial t}\right \} .  \label{IEpumpedsecondorder}
\end{equation}%
As a consequence of unitarity of the $S$-matrix (see Appendix~\ref%
{consunitarity}), the first term vanishes to second order in the precession
frequency: 
\begin{equation}
I_{\text{out}}^{(e)}=\frac{\hbar }{4\pi }\int_{0}^{\tau }\frac{dt}{\tau }%
\int d\epsilon \left( -\frac{\partial f}{\partial \epsilon }\right) \text{%
Tr}\left \{ \frac{\partial S_{\text{fr}}}{\partial t}\frac{\partial S_{%
\text{fr}}^{\dagger }}{\partial t}\right \} ,  \label{IEpumpedFinal}
\end{equation}%
where, \textit{at this point}, we may insert the frozen scattering matrix
since the current expression is already proportional to the square of the
pumping frequency. Furthermore, since there is no net pumped charge current
in one cycle (and we are assuming reservoirs in a common equilibrium), the
pumped heat current is identical to the pumped energy current, $I_{\text{%
out}}^{(q)}=I_{\text{out}}^{(e)}$.

Our expression for the pumped energy current (\ref{IEpumpedFinal}) agrees
with that derived in Ref.~\onlinecite{Moskalets:prb02} at zero temperature.
Our result (\ref{IEpumpedFinal}) differs from Ref. \onlinecite{Wang:prb02}
at finite temperatures. The discrepancy can be explained as follows.
Integration by parts over time $t$ in Eq. (\ref{IEsimplified}), using 
\begin{widetext}
\begin{equation}
\left[ f\left(\epsilon -\frac{i\hbar }{2}\frac{\partial }{\partial t }\right)i\hbar 
\frac{\partial S}{\partial t }\right] S^{\dagger }=2\left[ \epsilon
f\left(\epsilon -\frac{i\hbar }{2}\frac{\partial }{\partial t }\right)S\right]
S^{\dagger }-2\left[ \left(\epsilon -\frac{i\hbar }{2}\frac{\partial }{\partial
t }\right)f\left(\epsilon -\frac{i\hbar }{2}\frac{\partial }{\partial t }\right)S\right]
S^{\dagger },
\end{equation}%
and the unitarity condition from Appendix~\ref{consunitarity}, 
\begin{equation}
\int_{0}^{\tau} \frac{dt}{\tau} \int d\epsilon \left[ \left(\epsilon -\frac{i\hbar }{%
2}\frac{\partial }{\partial t }\right)f\left(\epsilon -\frac{i\hbar }{2}\frac{%
\partial }{\partial t }\right)S\right] S^{\dagger }=
\int_{0}^{\tau} \frac{dt}{\tau} \int d\epsilon \epsilon f(\epsilon ),
\end{equation}%
the DC pumped energy current can be rewritten as 
\begin{equation}
I_{\text{out}}^{(\epsilon )}=\frac{1}{\pi \hbar }
\int_{0}^{\tau} \frac{dt}{\tau} \int d\epsilon \text{Tr}\left \{ \left[ \epsilon f\left(\epsilon -%
\frac{i\hbar }{2}\frac{\partial }{\partial t }\right)S\right] S^{\dagger
}-\epsilon f(\epsilon )\right \}.
\end{equation}%
Next, we expand this to the second order in the pumping frequency and find 
\begin{equation}
I_{\text{out}}^{(\epsilon )}=\frac{1}{\pi \hbar }
\int_{0}^{\tau} \frac{dt}{\tau} \int d\epsilon \text{Tr}\left \{ \epsilon f(\epsilon )\left(
SS^{\dagger }-1\right) -\epsilon (\partial_\epsilon f)\frac{i\hbar }{2}\frac{\partial S%
}{\partial t }S^{\dagger }-\epsilon (\partial_\epsilon^2f)\frac{\hbar ^{2}}{8}%
\frac{\partial ^{2}S}{\partial t ^{2}}S^{\dagger }\right \} .
\label{IEanotherrep}
\end{equation}%
\end{widetext}This form of the pumped energy current, Eq.~(\ref{IEanotherrep}%
), agrees with Eq.~(10) in Ref.~\onlinecite{Wang:prb02} if one (\textit{%
incorrectly}) assumes $SS^{\dagger }=1$. Although for the frozen scattering
matrix, $S_{\text{fr}}S_{\text{fr}}^{\dagger }=1$, unitarity does not
hold for the Wigner representation of the scattering matrix to the second
order in the pumping frequency. $(SS^{\dagger }-1)$ therefore does not
vanish but contributes to leading order in the frequency to the pumped
current, which may not be neglected at finite temperatures. Only when this
term is included our new result Eq. (\ref{IEpumpedFinal}) is recovered.

\section{Fourier transform and Wigner representation}

There is a long tradition in quantum theory to transform the two-time
dependence of two-operator correlation functions such as scattering matrices
by a mixed (Wigner) representation consisting of a Fourier transform over
the time difference and an average time, which has distinct advantages when
the scattering potential varies slowly in time.\cite{Rammer:rmp85} In order
to establish conventions and notations, we present here a short exposure how
this works in our case.

The Fourier transform of the time dependent annihilation operators are
defined in Eqs.\ (\ref{cFourier1}) and (\ref{cFourier2}). Consider a
function $A$ that depends on two times $t_{1}$ and $t_{2}$, $%
A=A(t_{1},t_{2}) $. The Wigner representation with $t=(t_{1}+t_{2})/2$ and $%
t^{\prime }=t_{1}-t_{2}$ is defined as: 
\begin{align}
A(t_{1},t_{2})& =\frac{1}{2\pi \hbar }\int_{-\infty }^{\infty }d\epsilon
A\left( t,\epsilon \right) e^{-i\epsilon (t_{1}-t_{2})/\hbar },
\label{WignerRep} \\
A(t,\epsilon )& =\int_{-\infty }^{\infty }dt^{\prime }A\left( t+\frac{%
t^{\prime }}{2},t-\frac{t^{\prime }}{2}\right) e^{i\epsilon t^{\prime
}/\hbar }.
\end{align}%
We also need the Wigner representation of convolutions, 
\begin{equation}
(A\otimes B)(t_{1},t_{2})=\int_{-\infty }^{\infty }dt^{\prime
}A(t_{1},t^{\prime })B(t^{\prime },t_{2}).
\end{equation}%
By a series expansion, this can be expressed as\cite{Rammer:rmp85}%
\begin{equation}
(A\otimes B)(t,\epsilon )=e^{-i\left( \partial _{t}^{A}\partial _{\epsilon
}^{B}-\partial _{t}^{B}\partial _{\epsilon }^{A}\right) /2}A(t,\epsilon
)B(t,\epsilon )  \label{Wignerconvolution}
\end{equation}%
which we use in the following section.

\section{Properties of $S$-matrix}

\label{consunitarity}

Here we discuss some general properties of the two-point time-dependent
scattering matrix. Current conservation is reflected by the unitarity of the 
$S$-matrix which can be expressed as 
\begin{widetext}
\begin{equation}
\sum_{\beta n^{\prime }s^{\prime }}\int dt^{\prime }S_{n_{1}s_{1},n^{\prime
}s^{\prime }}^{(\alpha _{1}\beta )}(t_{1},t^{\prime
})S_{n_{2}s_{2},n^{\prime }s^{\prime }}^{(\alpha _{2}\beta )\ast }(t^{\prime
},t_{2})=\delta _{n_{1}n_{2}}\delta _{s_{1}s_{2}}\delta _{\alpha
_{1}\alpha _{2}}\delta (t_{1}-t_{2}).
\end{equation}
Physically, this means that a particle entering the scattering region from a lead $\alpha$ at some time $t$ is bound to exit 
the scattering region in some lead $\beta$ at another (later) time $t'$.
Using Wigner representation (\ref{WignerRep}) and integrating over the
local time variable, this implies (using Eq.\ (\ref{Wignerconvolution}))
\begin{equation}
1=\left( S \otimes S^{\dag }\right) (t ,\epsilon )  = 
e^{-i%
\left( \partial _{t }^{S}\partial _{\epsilon }^{S^{\dag }}-\partial
_{t }^{S^{\dag }}\partial _{\epsilon }^{S}\right)/2} S(t
,\epsilon )S^{\dag }(t ,\epsilon ),  \label{unitaritySSdagWigner}
\end{equation}%
where $1$ is a unit matrix in the space spanned by the wave guide modes (labelled by spin $s$ and orbital quantum number $n$).
Similary, we find
\begin{equation}
1=\left( S^{\dag }\otimes S\right) (t ,\epsilon )=e^{ +i%
\left( \partial _{t }^{S}\partial _{\epsilon }^{S^{\dag }}-\partial
_{t }^{S^{\dag }}\partial _{\epsilon }^{S}\right)/2} S^{\dag }(t
,\epsilon )S(t ,\epsilon ).  \label{unitaritySdagSWigner}
\end{equation}%
To second order in the precession frequency, by respectively subtracting and summing Eqs.~(\ref%
{unitaritySSdagWigner}) and (\ref{unitaritySdagSWigner}) give
\begin{equation}
\text{Tr}\left \{ \frac{\partial S}{\partial t }\frac{\partial S^{\dag }}{%
\partial \epsilon }-\frac{\partial S}{\partial \epsilon }\frac{\partial
S^{\dag }}{\partial t }\right \} =0  \label{Poissonzero}
\end{equation}%
and 
\begin{equation}
\text{Tr}\left \{ SS^{\dag }-1 \right \} =\text{Tr}\left \{ \frac{%
\partial ^{2}S}{\partial t ^{2}}\frac{\partial ^{2}S^{\dag }}{\partial
\epsilon ^{2}}-2\frac{\partial ^{2}S}{\partial t \partial \epsilon }\frac{%
\partial ^{2}S^{\dag }}{\partial t \partial \epsilon }+\frac{\partial
^{2}S}{\partial \epsilon ^{2}}\frac{\partial ^{2}S^{\dag }}{\partial t
^{2}}\right \} .
\end{equation}%
Furthermore, for any energy dependent function $Z(\epsilon )$ and arbitrary
matrix in the space spanned by spin and transverse waveguide modes $Y$, Eq.~(\ref%
{unitaritySSdagWigner}) implies
\begin{equation}
\frac{1}{\tau}\int_{0}^{\tau}dt\int d\epsilon Z(\epsilon )\text{Tr}\left \{ \left[
e^{-i\left( \partial _{t}^{S}\partial _{\epsilon
}^{S^{\dag }}-\partial _{t}^{S^{\dag }}\partial _{\epsilon }^{S}\right)/2} %
S(t,\epsilon )S^{\dag }(t,\epsilon )-1\right] Y\right \} =0.
\end{equation}%
Integration by parts with respect to $t$ and $\epsilon $ gives%
\begin{equation}
\frac{1}{\tau}\int_{0}^{\tau}dt\int d\epsilon \text{Tr}\left \{ \left[ e^{ -%
i\left( \partial _{t}^{S}\partial _{\epsilon }^{S^{\dag
}}-\partial _{t}^{S}\partial _{\epsilon }^{ZS^{\dag }}\right)/2}
S(t,\epsilon )Z(\epsilon )S^{\dag }(t,\epsilon )-Z(\epsilon )\right]
Y\right \} =0,
\end{equation}%
which can be simplified to 
\begin{equation}
\frac{1}{\tau}\int_{0}^{\tau}dt\int d\epsilon \text{Tr}\left \{ \left( \left[
Z\left(\epsilon +\frac{i}{2}\frac{\partial }{\partial t}\right)S(t,\epsilon )\right]
S^{\dag }(t,\epsilon )-Z(\epsilon )\right) Y\right \} =0.
\label{unitSSdagcondition}
\end{equation}%
Similarly from (\ref{unitaritySdagSWigner}), we find%
\begin{equation}
\frac{1}{\tau}\int_{0}^{\tau}dt\int d\epsilon \text{Tr}\left \{ \left( S^{\dag
}(t,\epsilon )\left[ Z\left(\epsilon -\frac{i}{2}\frac{\partial }{\partial t}%
\right)S(t,\epsilon )\right] -1\right) Y\right \} =0.  \label{unitSdagScondition}
\end{equation}
\end{widetext}Using this result for $Y=1$ and $Z(\epsilon )=f(\epsilon )$ in
the expression for the DC particle current (\ref{DCeqpumpedcurrent}), we see
that unitarity indeed implies particle current conservation, 
$\sum_{\alpha }I_{\alpha }^{(c)}=0$ for a time-periodic potential. However,
such a relation does not hold for spins. Choosing $Y=\boldsymbol{\sigma }$,
we cannot rewrite Eq.~(\ref{unitSdagScondition}) in the form (\ref%
{DCeqpumpedspincurrent}), unless the $S$-matrix and the Pauli matrices
commute. Unless the $S$-matrix is time or spin independent, a net spin
current can be pumped out of the system, simultaneously exerting a torque on
the scattering region.

Furthermore, choosing $Z(\epsilon )=\int_{0}^{\epsilon }d\epsilon ^{\prime
}f(\epsilon ^{\prime })$, $Y=1$ and expanding the difference between (\ref%
{unitSdagScondition}) and (\ref{unitSSdagcondition}) to second order in
frequency gives 
\begin{equation*}
\frac{1}{\tau }\int_{0}^{\tau }dt\int d\epsilon \text{Tr}\left \{
f(\epsilon )\frac{\partial S(t,\epsilon )}{\partial t}S^{\dag }(t,\epsilon
)\right \} =0,
\end{equation*}%
which we use to eliminate the first term in the expression for the energy
pumping, Eq.~(\ref{IEpumpedsecondorder}).


\end{document}